# Aging Effects on Superconducting Properties of BiS$_2$-Based Compounds: First-12-Year Restudy

Poonam Rani[1,2], Rajveer Jha[1,2,3], V. P. S. Awana[2], Yoshikazu Mizuguchi[1,*]

[1]Department of Physics, Tokyo Metropolitan University, Hachioji 192-0397, Japan.
[2]CSIR-National Physical Laboratory, New Delhi 110012, India.
[3]Microelectronics Research Center, The University of Texas at Austin, Austin TX 78758, USA.

**Abstract**

Decomposition of superconductors sometimes becomes crucial when studying essential physical properties of the superconductors. For example, the cuprate superconductor YBa$_2$Cu$_3$O$_{7-d}$ decomposes by long-time air exposure. In this study, we investigate the aging effects on superconducting properties of BiS$_2$-based superconductors Bi$_4$O$_4$S$_3$ and LaO$_{0.5}$F$_{0.5}$BiS$_2$, both were first synthesized in 2012, using their polycrystalline samples synthesized several years ago. We find that 12-year-old Bi$_4$O$_4$S$_3$ samples exhibit bulk superconductivity with a slight degradation of the superconducting transition temperature ($T_c$) of 0.2 K. For a high-pressure-synthesized LaO$_{0.5}$F$_{0.5}$BiS$_2$ sample, clear decrease in $T_c$ is observed, which suggests that high-pressure strain is reduced by aging.

**Keywords:** BiS$_2$-based superconductor, magnetization, aging effect, air exposure

1. **Introduction**

Superconductivity is a quantum phenomenon characterized with zero-resistivity states, diamagnetism, and Josephson effects, and the superconducting states emerge at temperatures below the transition temperature ($T_c$). Using those properties, superconductivity application has been developed and will be more diverse in near future [1]. Since application of superconducting devices requires stability of the components of the device, degradation of the properties and/or decomposition of the compound components become crucial problems. For example, cuprate (Cu-oxide) high-$T_c$ superconductor $YBa_2Cu_3O_{7-d}$ decomposes by long-time air exposure [2-4]. In addition, the use of reactive ions in the constituent elements or the presence of quasi-stable phases with a formation energy comparable to the target phase results in decomposition of the obtained samples by aging in the air-exposure environment [5-7]. Knowledge on phase stability and aging effects is also important when studying superconducting properties because the change in superconducting properties after aging affects the conclusion of the research. For some rare cases, positive aging effects are observed, and one of the examples is the improvement of $T_c$ in Fe(Te,S) by aging with air exposure [8].

Here, we investigate the aging effects of $BiS_2$-based superconductors $Bi_4O_4S_3$ and $LaO_{0.5}F_{0.5}BiS_2$. Those superconductors were discovered in 2012 [9-11], followed by the discovery of various related superconductors [12-22]. The crystal structure of $BiS_2$-based compounds is basically composed of alternate stacks of a blocking (insulating) layer and a $BiS_2$ conducting layer (Fig. 1), which is similar to stacking structures of cuprates and Fe-based high-$T_c$ superconductors [23-26]. The $Bi_4O_4S_3$ sample exhibits bulk superconductivity without any high-pressure treatments [9,27-30]. In contrast, the polycrystalline sample of $LaO_{0.5}F_{0.5}BiS_2$ does not exhibit bulk superconductivity, and high-pressure annealing or synthesis under high-pressure conditions (with 1-3 GPa) is needed to induce bulk superconductivity [10,31-35]. Furthermore, degradation of superconducting properties of ambient-pressure annealing of the high-pressure-annealed (bulk-superconducting) $LaO_{0.5}F_{0.5}BiS_2$ sample was observed [36]. The cause of the absence of bulk superconductivity in La(O,F)$BiS_2$ systems are explained by local Bi-S distance disorder [37-42] and the formation of superstructure [43]. Therefore, the high-pressure effects in the $LaO_{0.5}F_{0.5}BiS_2$ sample should be affected by aging even at ambient temperature condition, and studying aging effects after keeping the samples for long time at ambient conditions enables us to know their aging effects. Here, we re-investigate the superconducting properties of the oldest $Bi_4O_4S_3$ sample synthesized in April of 2012 [9,27] and the $Bi_4O_4S_3$ $^{34}S$-isotope samples prepared in 2020 [44]. In addition, the $LaO_{0.5}F_{0.5}BiS_2$ sample high-pressure-synthesized in 2014 [35] is studied. In Japan, the birth years are categorized with Japanese zodiac, and people use the 12-year period to celebrate anniversaries. This year is the first 12 years since the discovery of $BiS_2$-based superconductors, and reinvestigations every 12 years will be useful.

2. Experimental

The polycrystalline samples of $Bi_4O_4S_3$ and $LaO_{0.5}F_{0.5}BiS_2$ were previously synthesized and stored in a glass container with air in an office of Tokyo Metropolitan university (TMU, Hachioji city). The details of sample preparation methods are reported in Refs. 9, 12, 20, and 29. For the $LaO_{0.5}F_{0.5}BiS_2$ sample, high-pressure sintering under 2 GPa at 700 °C was performed in 2014 [35]. Here, we label the $Bi_4O_4S_3$ samples as follows. $Bi_4O_4S_3$ (#1) was prepared by Mizuguchi et al. in 2012 at TMU [9]. $Bi_4O_4S_3$ (#2) was prepared by Singh et al. in 2012 at National Physical Laboratory (NPL)-India [27] and send to TMU, followed by keeping in TMU. $Bi_4O_4S_3$ (#3) with $^{34}S$ was synthesized in 2020 for an isotope-effect study [44].

Temperature ($T$) dependences of magnetization ($M$) after both zero-field cooling (ZFC) and field cooling (FC) were measured using a superconducting quantum interference device (SQUID) magnetometer with an applied magnetic field ($H$) of $H$ = 10 Oe. Powder X-ray powder diffraction (XRD) was carried out using RIGAKU diffractometer Miniflex-II (in 2012) and Miniflex 600 (in 2024).

3. Results and Discussion

Figure 2 shows the comparison of XRD patterns of $Bi_4O_4S_3$ (#1) taken in 2012 (as-synthesized) and 2024 (12-years old). Except for the differences in the intensity ratio, which are caused by the sample setting, the observed patterns are quite similar. Therefore, the $Bi_4O_4S_3$ phase does not exhibits clear decomposition in the level of laboratory XRD in 12 years. Figure 3 shows the $T$ dependances of $M$ for $Bi_4O_4S_3$ (#1) measured in (a) 2012 and (b) 2024. Bulk superconductivity signals are observed in both data, which is consistent with the XRD results. We estimated $T_c^{onset}$ from cross point of two lines as shown in the insets. $T_c^{onset}$ slightly decreased from 5.0 to 4.8 K in 12 years. This change might be caused by slight change in crystal structure or carrier concentration because the $Bi_4O_4S_3$ phase possesses a complicated stacking structure. Another potential reason for the $T_c$ change is release of strain effect. Although the $Bi_4O_4S_3$ (#1) sample was synthesized at ambient pressure, the as-synthesized sample might have some strain effects generated during the pelletizing and/or annealing processes. However, the change in $T_c$ is quite small, and we consider that there is almost no aging effect in the 12-years-old $Bi_4O_4S_3$ sample.

Next, we see the current superconducting property of $Bi_4O_4S_3$ (#2) synthesized in NPL-India and stored in TMU for more than 11 years. Figure 4 shows the $T$ dependances of $M$ for $Bi_4O_4S_3$ (#2) measured in 2024. Bulk superconductivity is confirmed by the large diamagnetic signals, and $T_c^{onset}$ is 4.4 K, which is comparable to $T_c$ reported in Ref. 27. In addition, $M$-$T$ for the $Bi_4O_4S_3$ (#3) sample with $^{34}S$ is shown in Fig. 5. As well known, $^{32}S$ is dominant in nature and $^{34}S$ is a rarer isotope that $^{32}S$. The comparison of ZFC data clearly shows that there is no degradation of superconducting property. The estimated $T_c$s are almost same, 4.81 K (taken in 2020) and 4.80 K (taken in 2024).

Finally, we investigate the superconducting property of $LaO_{0.5}F_{0.5}BiS_2$ synthesized by high-pressure sintering under 2 GPa (HP-$LaO_{0.5}F_{0.5}BiS_2$). Figure 6 shows the $T$ dependances of $M$ for HP-$LaO_{0.5}F_{0.5}BiS_2$ measured in (a) 2014 and (b) 2024. Clear decrease in $T_c$ is observed after 10-years storage. Although bulk nature of superconductivity is maintained, the irreversibility temperature ($T_{irr}$), which is estimated from the bifurcation point of ZFC and FC data, decreases from 9.4 to 7.0 K. Because $T_{irr}$ of the $BiS_2$-based samples roughly corresponds to zero-resistivity temperature ($T_c^{zero}$) in resistivity measurement, $T_c^{zero}$ should be degraded in 10 years. Because of the limitation of the original sample of HP-$LaO_{0.5}F_{0.5}BiS_2$, we cannot perform resistivity experiments. The aging effect, the decrease in superconducting property, in HP-$LaO_{0.5}F_{0.5}BiS_2$ would be explained by the reduction of strain effects generated by high-pressure sintering. As shown in Ref. 45, crystal structure changes from tetragonal to monoclinic by external pressure. The high-pressure-synthesized samples also exhibits XRD peak broadening even after removing pressure [9,34,35]. The broadened XRD peak is recovered by annealing the sample at ambient pressure [36]. Therefore, the observed aging effect should be related to the release of high-pressure strain effects in the sample.

## 4. Conclusion

We investigated the aging effects on crystal structural, decomposition, and superconducting properties of $BiS_2$-based compounds $Bi_4O_4S_3$ and $LaO_{0.5}F_{0.5}BiS_2$ using XRD and magnetization measurements. For the $Bi_4O_4S_3$ sample, no aging effect on XRD was observed, and quite small decrease in $T_c$ (0.2 K) was observed for the 12-years-old sample. As a conclusion, there is almost no aging effects in $Bi_4O_4S_3$. In contrast, the HP-$LaO_{0.5}F_{0.5}BiS_2$ sample exhibits a clear aging effect in magnetization data. $T_{irr}$ decreases from 9.4 to 7.0 K. Because the high-pressure-synthesized sample contain high-pressure strain effects, the aging effects on superconducting properties were observed in 10-years-old samples. As a conclusion, the $BiS_2$-based compounds are stable and their superconducting properties are also robust to aging in air. However, the high-pressure strain effects exhibit aging.

**Acknowledgements**

We thank all the coauthors of Refs. 9, 27, and 35. This work was partly supported by TMU research project for emergent future society.

*E-mail: mizugu@tmu.ac.jp

**Statements and Declarations**

The authors declare no competing interests.

Table 1: Sample information and $T_c$s.

| Sample | Ref. | $T_c$ (as-synthesized) | $T_c$ (2024) |
|---|---|---|---|
| $Bi_4O_4S_3$ (#1) | [9] | $T_c^{onset}$ = 5.0 K | $T_c^{onset}$ = 4.8 K |
| $Bi_4O_4S_3$ (#2) | [27] | $T_c^{onset}$ = 4.4 K [27] | $T_c^{onset}$ = 4.4 K |
| $Bi_4O_4{}^{34}S_3$ (#3) | [44] | $T_c^{onset}$ = 4.81 K | $T_c^{onset}$ = 4.80 K |
| HP-$LaO_{0.5}F_{0.5}BiS_2$ | [35] | $T_{irr}$ = 9.4 K | $T_{irr}$ = 7.0 K |

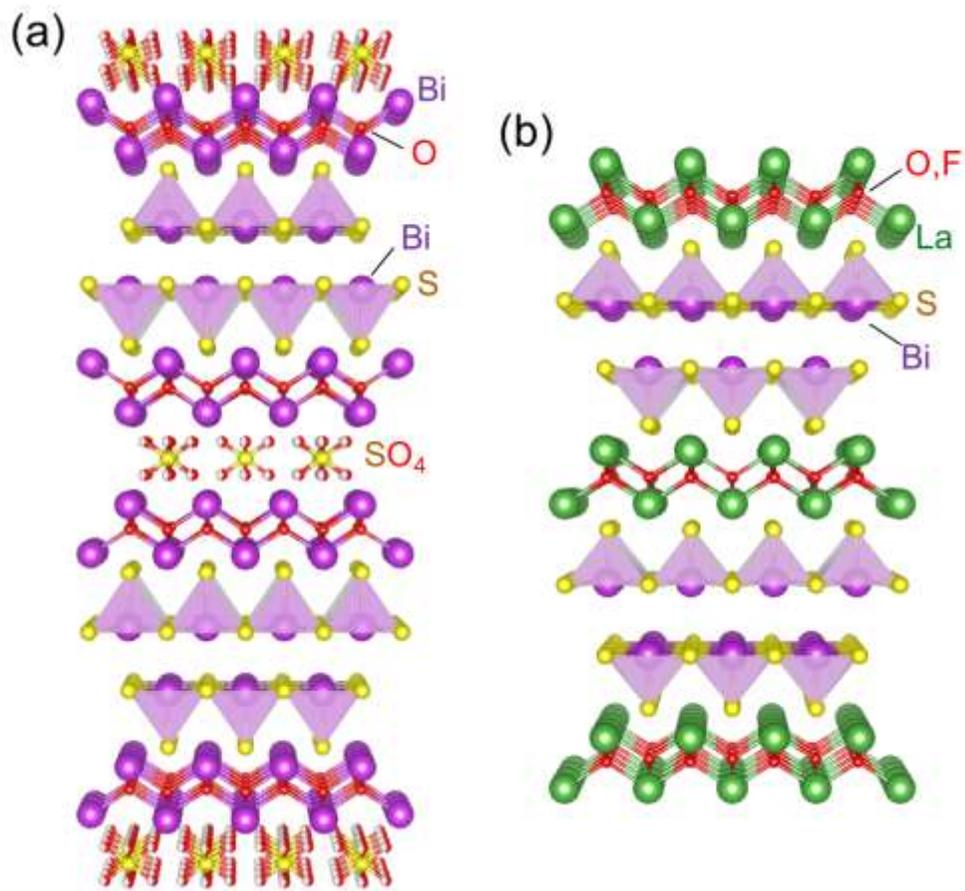

Fig. 1. Crystal structures of $Bi_4O_4S_3$ and $LaO_{0.5}F_{0.5}BiS_2$.

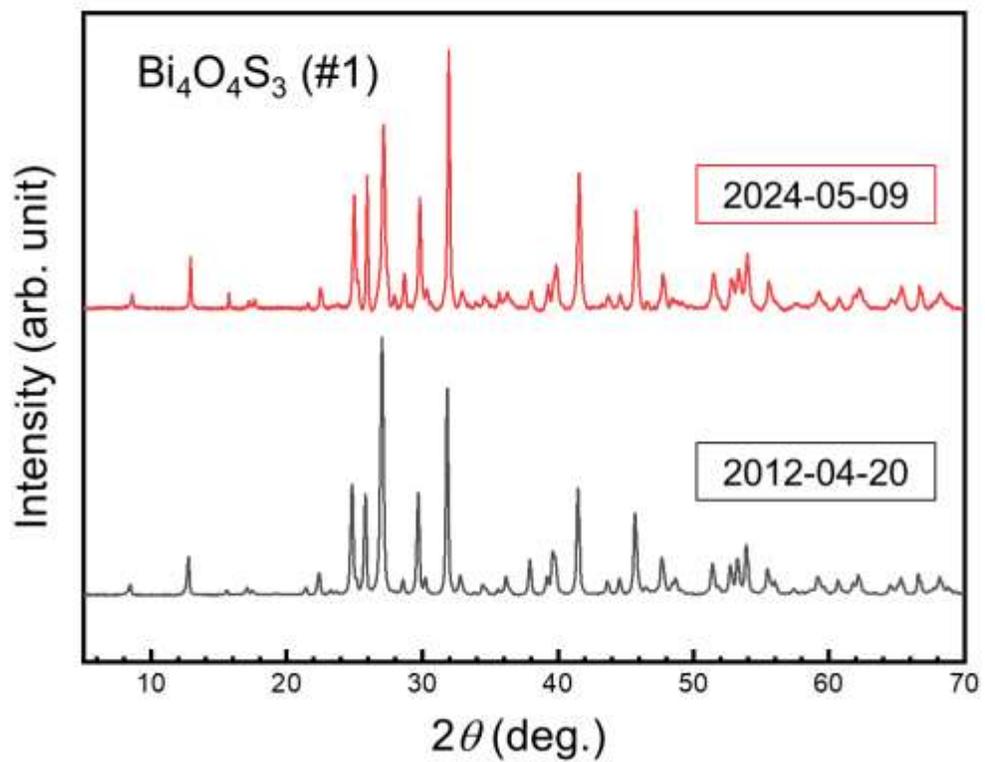

Fig.2. XRD patterns for $Bi_4O_4S_3$ (#1) measured in 2012 (2012-04-20) and 2024 (2024-05-09).

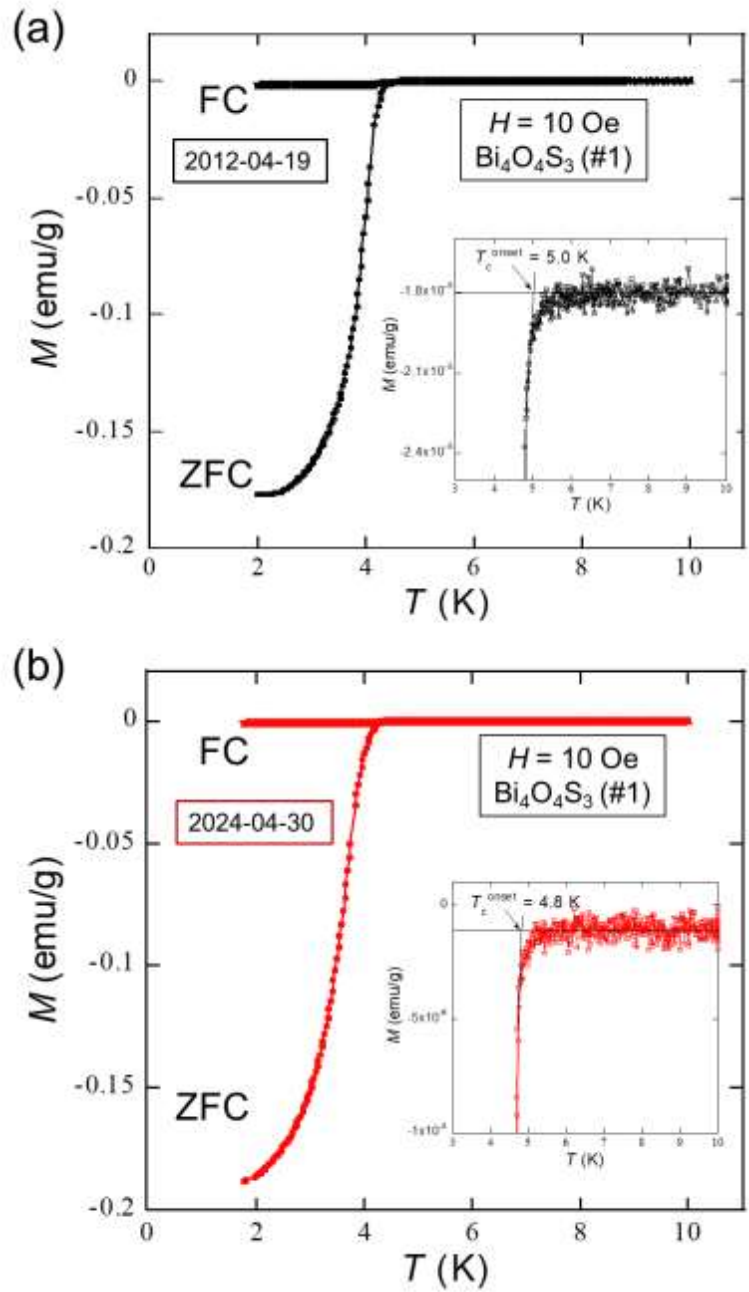

Fig.3. $T$ dependences of $M$ for $Bi_4O_4S_3$ (#1) measured in (a) 2012 and (b) 2024. The insets show data near $T_c^{onset}$.

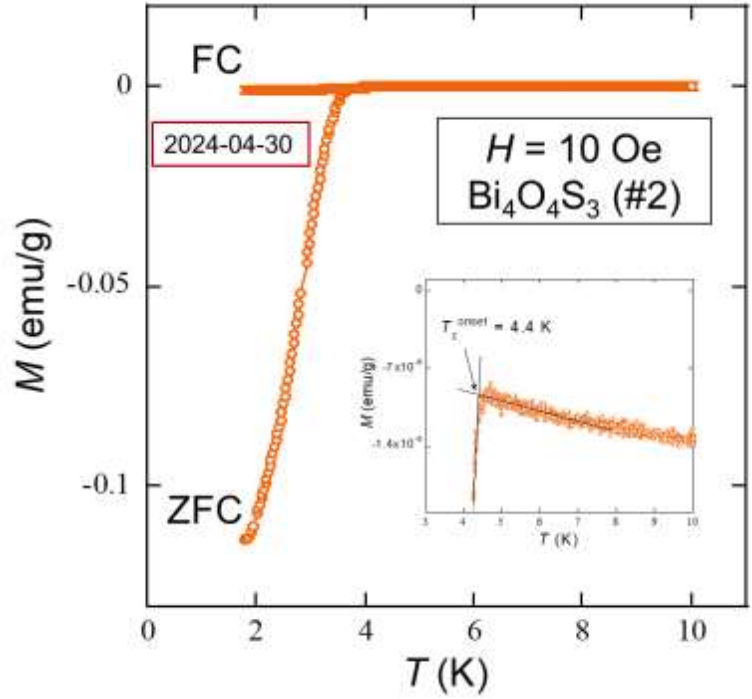

Fig.4. $T$ dependence of $M$ for $Bi_4O_4S_3$ (#2) measured in 2024. The inset shows data near $T_c^{onset}$.

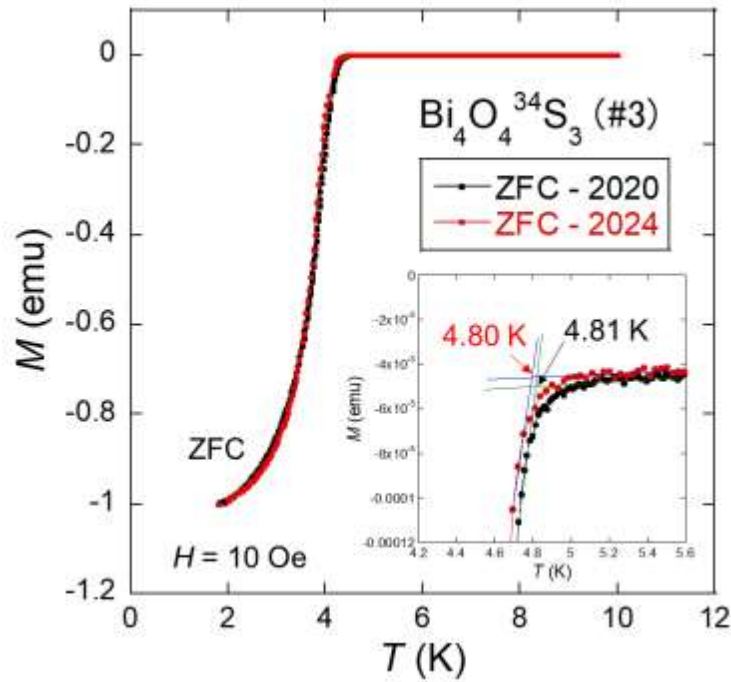

Fig.5. $T$ dependences of $M$ (ZFC) for $Bi_4O_4S_3$ (#3) measured in 2020 and 2024. The inset shows data near $T_c^{onset}$.

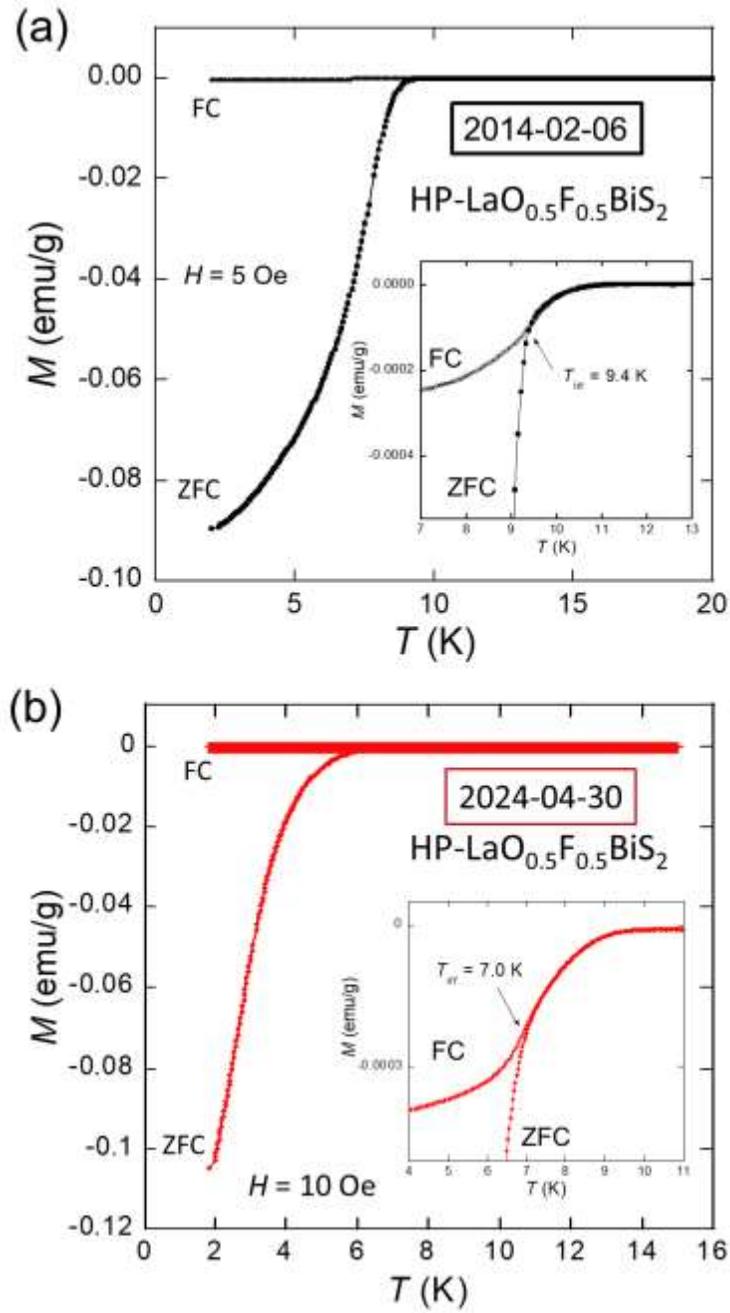

Fig.6. $T$ dependences of $M$ for HP-LaO$_{0.5}$F$_{0.5}$BiS$_2$ synthesized at 2 GPa measured in (a) 2014 and (b) 2024. The insets show data near $T_{\mathrm{irr}}$.